# INSTANTANEOUS NOISE-BASED LOGIC [a]

LASZLO B. KISH [1], SUNIL KHATRI [1] AND FERDINAND PEPER [2]

[1] *Department of Electrical and Computer Engineering, Texas A&M University, College Station, TX 77843-3128, USA*

[2] *National Institute of Information and Communications Technology, Kobe, 651-2492 Japan*

We show two universal, Boolean, deterministic logic schemes based on binary noise timefunctions that can be realized without time-averaging units. The first scheme is based on a new bipolar random telegraph wave scheme and the second one makes use of the recent noise-based logic which is conjectured to be the brain's method of logic operations [Physics Letters A 373 (2009) 2338-2342]. Error propagation and error removal issues are also addressed.

## 1. Introduction

Recently, a new type of logic has been proposed in which the logic 0 and logic 1 signals are represented – not by discrete electrical values like in current computers – but by independent noise sources [1,2]. A sinusoidal realization of this concept is also possible [3] but it is out of the scope of the present paper. Since these noise sources are statistically independent, their correlation, averaged over time, is zero, and this forms the basis for defining these sources as orthogonal. In order to produce the output of a circuit, the input information is first interpreted by correlating the input logic signal with the reference noises. For binary signals the independent random signals $H(t)$ and $L(t)$ are used to encode the binary values 1 and 0, respectively [1]. A signal input to a circuit, say $X(t)$, is encoded in terms of the reference signals, so, $X(t) = L(t)$ or $X(t) = H(t)$. Multiplying $X(t)$ by a reference noise signal (either by $L(t)$ or $H(t)$) and averaging the resulting product over time, we obtain a value that is near 0 in case $X(t)$ is different from the actual reference signal, or much higher otherwise. This time-average is then used in a logic operation that additionally takes $X(t)$ and/or the reference noise signals as an input, to produce the output. Gates in noise-based logic can be implemented using simple analog circuits like analog multipliers, RC-circuits for time-averaging, and analog switches to implement the output controlled by the time-averaged input values. This concept has been generalized to multi-valued logic computations as well [2].

---

[a] http://arxiv.org/abs/1004.2652

*Instantaneous noise-based logic*The $L(t)$ and $H(t)$ noises are orthogonal not only each other but also to any other noise sources in the system, such as the background noise of the devices, thus the impact of such noise sources will be automatically removed in the averaging process [1]. Moreover, it was shown that the switches can be of poor quality with high error rate and the logic can still operate with the low error rates known for conventional digital circuitry. Thus, in principle, a very low supply voltage can be used which results in a reduced power dissipation [1].

There is also another new potential: noise-based logic allows – in principle – the use of superpositions of different signals [1]. Moreover, high-dimensional logic spaces can be defined [1,2] similarly to quantum computation. This holds the potential of being able to achieve multi-valued logic, along with significant parallelism based on such superpositions of hyperspace elements, which may result in increased computation speed or hardware efficiency. As an illustration of the power of the superposition of high-dimensional hyperspace elements, in [2] a string search algorithm was shown with the same hardware complexity as Grover's quantum search engine but the number of required operations are reduced by a factor of $\sqrt{N}$, where $N$ is the length of the strings.

The noise-based logic schemes proposed thus far, however, suffer from a significant slow-down due to the need for a time-averaging process, which is conducted in every gate. To get a low error probability in the order of $10^{-25}$, the clock frequency will be about 600 times lower [1] than the small-signal bandwidth of the circuitry, see also Section 5. That means, these classical noise-based logic schemes [1,2] must operate at 0.5 GHz clock frequency or less with the current microtechnology [1].

Alternative schemes are thus needed that avoid the time-averaging process. We refer to these schemes as *Instantaneous noise based logic* (*INBL*) schemes. Such schemes would result in drastically increased computation speeds (by an order of magnitude or two), and thus allow the full potential of noise-based logic to be realized. Averaging would still be needed at the output interfaces, where conventional noise-based logic [1,2] should be used, however that represents only a minor slowdown provided significantly complex combinational operations are done before that in the logic, without the need of time-averaging.

It is interesting to note that the noise-based brain logic scheme proposed recently [5] is also an *INBL* scheme.

This paper proposes two noise-based Boolean logic schemes that do not require a time-averaging process. We prove that these logics are universal and we show some of the related gates. The first logic is based on bipolar random telegraph waves. The second approach utilizes the logic scheme and basic operations of [5] which we will use to introduce Boolean logic gates there.

In this paper, we work with squeezed (called degenerated in [2]) logic states, which means that the $H$ logic value of a given bit is represented by a noise $H(t)$ that is orthogonal to the noise representing the $H$ value of other binary digits but the $L$ logic value is represented by "no signal" or $L(t) = 0$ for all binary digits. Thus the $L$ value is common for all the bits. We will define the Boolean logic values and the NOT and AND gates. Hence the universality of the logic is proven.

Finally, we will comment on higher dimensional multivalued logic vectors (hyperspace).

## 2. Random telegraph wave based logic

Let us suppose that the noise functions $H(t)$ representing the logic value $H$ is a random telegraph wave (RTW) defined as follows. The absolute value of the amplitude is 1, and at the beginning of each time step, the amplitude changes its sign with probability 1/2. Thus the RTW wave is a square wave with 50% probability to have a +1 value, and the same probability to have a -1 amplitude.

The NOT gate operation is defined by Equation 1, and its hardware is a linear differential amplifier shown in Figure 1.

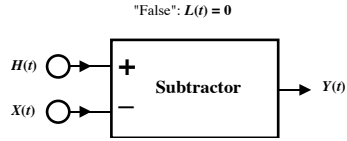

**Figure 1**. The NOT (INVERTER) gate. It consists of a linear amplifier.

$$Y(t) = \text{NOT } X(t) = H(t) - X(t) \tag{1}$$

If $X(t) = L(t) = 0$ then $Y(t) = H(t)$. If $X(t) = H(t)$, then $Y(t) = 0 = L(t)$.

The AND gate operation is defined by Equation 2 and it's hardware consists of two multipliers as shown in Figure 2:

$$Y(t) = X_1(t) \text{ AND } X_2(t) = X_1(t) X_2(t) H(t) \tag{2}$$

The output will produce $H(t)$ when both inputs have $H(t)$ signal. Otherwise the output produces $L(t)$. Note that the multipliers can be simplified to linear amplifiers (with amplification of +1 and -1, respectively) and switches because of the trivial tasks of multiplication by ±1.

*Instantaneous noise-based logic*

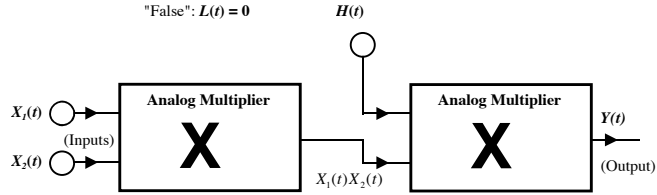

**Figure 2**. The AND gate.

Since both the AND and NOT gates can be realized, we have proved that the RTW-based *INBL* scheme defined above is universal.

## 3. Random spike train based logic

In [5], a model for instantaneous, deterministic, multi-valued logic was presented, and was conjectured to be utilized by the brain. In this logic, set-theoretical operations on randomly occurring, non-overlapping uniform and unipolar neural spike trains were used to construct logic superpositions and to analyze them. The key neural circuit element was the *orthon*, see Figure 3, which consisted of two neurons. The idealized neurons (see Figure 3, right hand side) have two inputs, an excitatory (+) and an inhibitory (-) one. Pulses arriving at (+) will propagate to the output of the neuron unless, at the same time, a pulse arrives also at the (-) input. Note, these idealistic neurons are supposed to be free of delays [5]. The orthon has two inputs and the *A* and *B* spike trains are treated as sets of spikes. The two outputs [5] compute the set-theoretical $AB$ ( $A \cap B$ ) operation, where the overlapping spikes of *A* and *B* are kept and the rest are discarded; and the set theoretical $A\overline{B}$ ( $A \cap \overline{B}$ ) operation, where the spikes of *A* not overlapping with *B* are kept and the rest are discarded).

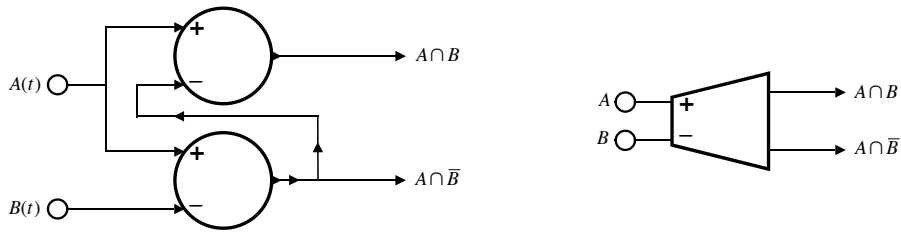

**Figure 3**. Left: the neural circuit of the *orthon* [5]. Right: its symbol.

We now show the universality of the relevant (squeezed) binary logic, when the spike train representing the logic value $H$ is the spike train $H(t) \neq 0$ and the logic value $L$ is represented by no spikes ( $L(t) = 0$ ). The NOT gate is defined as:

$$Y(t) = \text{NOT } X(t) = H(t) \cap \overline{X}(t) \tag{3}$$

For $X(t) = H(t)$, $Y(t) = L(t) = 0$ and for $X(t) = L(t) = 0$, $Y(t) = H(t)$.

The AND gate is defined as:

$$Y(t) = X_1(t) \text{ AND } X_2(t) = X_1(t) \cap X_2(t) \tag{4}$$

The orthon-based representation of the NOT and AND gates are shown in Figure 4. An orthon can be used to perform both logic functions.

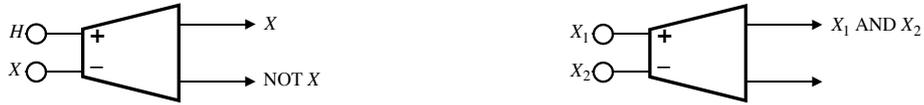

**Figure 4**. The binary NOT (left) and AND (right) gates utilize the orthon element of spike-based logic. Note that the upper output is not used in the circuit on the left, and the lower output is not used in the circuit on the right.

In this way we proved that the spike-based *INBL* scheme defined above is universal as well.

Both the RTW and the random spike train based approaches are *INBL* schemes. This is because the output can be determined to be logic $H$ or $L$ after the first pulse period

## 4. On multivalued logic and the hyperspace

To keep the RTW-based logic instantaneous, additive superpositions are not allowed because time averages to analyze superpositions are not permitted. However, multivalued logic can still be realized by building hyperspace product vectors [1,2] by multiplying the RTWs representing the non-zero bit values:

$$Z(t) = \prod_i H_i(t) \tag{5}$$

where the *i* index runs over those noise bits only which are present in the hyperspace element. With *n* RTWs, we can generate $2^n - 1$ hyperspace elements, thereby realizing multi-valued logic signals. It is interesting to note that the statistical properties and allowed values of $Z(t)$ are the same as that of any of the $H_i(t)$ RTWs. Moreover:

$$H_i^{2p}(t) = 1 \qquad \text{and} \qquad H_i^{2p-1}(t) = H_i(t), \tag{6}$$

where $p$ is and integer number.

So, if we want to remove the $H_k(t)$ element from the hyperspace product, we multiply the hyperspace vector by it:

$$Z'(t) = H_k(t)Z(t) = H_k(t)\prod_i H_i(t) = \prod_{i \neq k} H_i(t) \tag{7}$$

If we want to put the $H_k(t)$ element back into the hyperspace product, we apply the operation again:

$$Z(t) = H_k(t)Z'(t) = H_k(t)\prod_{i \neq k} H_i(t) = \prod_i H_i(t) \tag{8}$$

Interestingly, the hyperspace construction for the random spike train based logic scheme remains the same as the already published version [5]: it can be synthesized and analyzed/modified by set theoretical operations That means simple addition of orthogonal spike trains can be used to construct hyperspace elements (by performing the union of orthogonal spike trains). Similarly, the orthon can be used to remove components or to analyze the superposition, see the Neural Fourier Transformation part of the paper [5].

## 5. On applications and errors

When power efficiency is an issue, the easiest way to act is to reduce the supply voltage of the logic circuitry to the minimum where stable operation still holds. However, the implication is a lower noise margin and an exponentially higher rate of *fast errors*. Here "fast" means a time scale of the correlation time of the background noise which is typically few percents of the clock period. Such a fast error easily becomes a slow or persistent error if it triggers a *flip-flop* or other memory device. From the instantaneous nature of *INBL* it follows that fast errors, false transients, spikes, etc, will quickly propagate to the output. It has close to 100% error propagation efficiency. This fact indicates limitations of direct applications, such as driving certain type of memories, etc. Before doing that, or before outputting the *INBL* to other logic systems, these fast errors must first be removed.

To remove such an errors, the classical noise-based logic [1] with its input correlators, time-average unit, and threshold operation (while driving its output switches [1]) is needed. If the incoming fast error probability $P_{fast} \ll 0.5$ then such errors can easily be removed by the classical noise-based logic gate but the price is slowing down [1]. In Figure 5, as an example, a possible solution to output a *INBL* value to a classical (DC) digital circuitry is shown by using the FOLLOWER gate of the classical noise-based logic. If the input signal is $H(t)$ then the output (DC) signal is the positive power supply voltage ($H$ value), otherwise the output is ground potential ($L$ value). The $RC$ time-average circuit removes the remaining noises at the output of the crosscorrelator and, for $P_{fast} << 0.5$, the DC threshold of the output switches provide correct operation even if their driving DC voltage slightly deviates from the idealistic value.

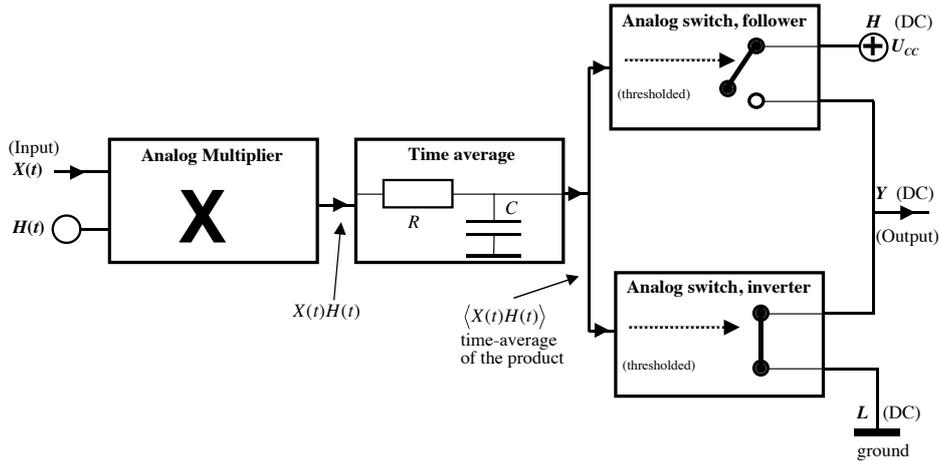

**Figure 5.** The FOLLOVER gate of the classical ("slow") noise-based logic used as interface toward regular (DC) digital circuitry. The time-averaging procedure, which is essential for the input correlator of the "slow" logic, together with the threshold of the output switches, removes the (fast) errors [1].

Alternatively, instead of using a FOLLOWER, the last logic gate in a dominantly *INBL* system, which realizes the required logic function, can be a classical, slow noise-based logic gate. In this case, only the last gate will be slow but rest of the system is *INBL* thus orders of magnitudes faster.

In this respect, there is some similarity with quantum computational engines. The classical, slow noise-based logic operation is like executing a quantum measurement. The *INBL* operations are like pure quantum operations without measurement. Fast errors are correspond to decoherence effects. A substantial difference compared to quantum computing is that the "measurement" by the classical noise-based logic gate is deterministic while quantum measurements are inherently statistical.

**6. Conclusion**

We have shown two binary, instantaneous noise-based logic (*INBL*) schemes and proved their universality. The first was a new, binary random telegraph wave (RTW) based system. The second one was the random spike train based logic system [5]. The latter scheme was shown to be inherently an *INBL*, and the only addition in this paper is the demonstration of universal binary gates for the latter scheme. These simple binary logics work with $L(t) = 0$ (squeezed logic).

The relevant hyperspace of the RTW-based logic was outlined and that of the spike-based logic is the same as earlier [5].

These logic systems have the potential to do fast noise-based logic operations. New applications of the multivalue logic represented by the hyperspace are to be studied. One very recently discovered application is the efficient verification of strings via a slow communication channel [6].

## Acknowledgements

LBK appreciates discussions with Tamas Horvath, Sergey Bezrukov and Zoltan Gingl.